\newcommand{\eg}{e.\,g.\ }
\newcommand{\idest}{i.\,e.\ }
\documentclass[12pt,preprint]{aastex}

\shorttitle{Molecular opacities for metal-poor AGB stars}
\shortauthors{Cristallo et al.}

\begin{document}

\title{Molecular opacities for low-mass metal-poor AGB stars undergoing the Third Dredge Up}

\author{S. Cristallo and O. Straniero}
\affil{INAF-Osservatorio Astronomico di Collurania,
    64100 Teramo, Italy}

\author{M.~T. Lederer and B. Aringer}
\affil{Institut f\"ur Astronomie, T\"urkenschanzstra\ss e 17,
A-1180 Wien, Austria}


\begin{abstract}

The concomitant overabundances of C, N and s-process elements are
commonly ascribed to the complex interplay of nucleosynthesis,
mixing and mass loss taking place in Asymptotic Giant Branch
stars. At low metallicity, the enhancement of C and/or N may be up
to 1000 times larger than the original iron content and
significantly affects the stellar structure and its evolution. For
this reason, the interpretation of the already available and still
growing amount of data concerning C-rich metal-poor stars
belonging to our Galaxy as well as to dwarf spheroidal galaxies
would require reliable AGB stellar models for low and very low
metallicities. In this paper we address the question of
calculation and use of appropriate opacity coefficients, which
take into account the C enhancement caused by the third dredge up.
A possible N enhancement, caused by the cool bottom process or by
the engulfment of protons into the convective zone generated by a
thermal pulse and the subsequent huge third dredge up, is also
considered. Basing on up-to-date stellar models, we illustrate the
changes induced by the use of these opacity on the physical and
chemical properties expected for these stars.

\end{abstract}


\section{Introduction}

The modification of the surface composition of a star may be due
to several processes. If the synthesis of nuclei occurs in stellar
interiors, the products of nuclear reactions only appear at the
surface when this chemically enriched material is moved from the
deepest layers to the external zone. The most common deep-mixing
processes are caused by convective instabilities. Less studied,
but not less important, are dynamical instabilities induced by
rotation or by magnetic forces \citep[see e.g.][]{mame}. Minor
surface chemical alterations occur on a longer timescale, as those
due to microscopic diffusion, levitation powered by radiation
pressure or other thermal instabilities (\citealp{kiwe},
\citealp[see also][and references therein]{pi07}). The first
systematic modification of the surface composition takes place at
the base of the red giant branch, when the external convection
penetrates the zone previously exposed to the H burning. This is
the so called {\it first dredge up} (FDU). Later on, during the
early Asymptotic Giant Branch, intermediate mass stars
($4<$M/M$_\odot<8$) eventually experience a {\it second dredge
up}. The main effect on the surface composition of these two
dredge-up episodes is the increase of the He abundance. A
redistribution of the CNO isotopes also occurs, namely isotopes
with a slow proton capture  (e.g. $^{14}$N) become more abundant,
whereas those whose proton capture is fast are depleted (e.g.
$^{12}$C). In any case, after the first and the second dredge up
the global number of CNO isotopes is conserved. This is not the
case of the {\it Third Dredge Up} (TDU). Actually, TDU refers to
multiple dredge up episodes occurring during the late part of the
AGB. In these stars, recursive Thermal Pulses (TPs) are powered by
violent He ignitions, which take place at the base of the thin
He-rich zone located between the CO core and the H-rich envelope
(intershell). These violent He burning episodes cause the
intershell to become instable against convection and therefore to
mix the products of the 3-$\alpha$ burning throughout this region.
Shortly after each thermal pulse, the H-burning dies down (owing
to the expansion initiated to counterbalance the excess of energy
released by the 3-$\alpha$ reactions) and the external convection
may penetrate down to the H-exhausted region. Thus, the stellar
envelope is enriched with the ashes of the He burning, mainly
$^{12}$C. This carbon dredge up is of great importance for the
future AGB evolution. Indeed, the efficiency of the CNO cycle and
that of the radiative energy transfer are both depending on the
carbon abundance in the envelope. During the AGB, owing to the
carbon enhancement, the opacity increases and, in turn, the
temperature gradient becomes larger\footnote {In the external
convective zone of an AGB stars, the radiative flux significantly
contributes to the overall energy transport.}. In practice, since
as a consequence of the C dredge up the C/O becomes rapidly
greater than 1, the effective temperature decreases, the stellar
radius increases and the average mass loss rate increases, thus
eroding at a faster rate the envelope mass. On the other hand, the
growth in mass of the H-exhausted core, which is controlled by the
H burning, also depends on the amount of C (and N) in the
envelope. As shown by \citealp{stra03}~(2003), changes of the core
and the envelope masses affect all the fundamental properties of
AGB stars, such as the thermal pulse strength or the total amount
of mass that is dredged up.

In the more massive AGB stars (M$>$4 M$_\odot$), an additional
phenomenon should be considered. In this stars, indeed, the
temperature at the base of the convective envelope may become
large enough (T$>5\times 10^7$ K) for the activation of the CN
cycle. In this case, the C excess in the envelope is partially
converted into N. This process is known as {\it Hot Bottom
Burning} \citep{sugi,ib73}.

In principle, a stellar evolution code should account for the
variations of the envelope chemical composition. In practice, only
variations of the main constituents are usually considered. In
particular, low temperature radiative opacity tables are available
only for scaled solar composition, so that only changes of H and
He can be accounted for. Although the use of these tables
substantially underestimates the radiative opacity of the cool
atmosphere of an evolved AGB star eventually enriched in C and N,
they are commonly adopted by stellar modelers, who are often left
without any other alternative.

\citealp{Marigo}~(2002) made a first step towards a correct
description of the abundance changes in the calculation of opacity
coefficients, by estimating molecular concentrations through
dissociation equilibrium calculations. Although the results of
this work definitely demonstrate the importance of a correct
opacity treatment, its simplified approach suffers from some
drawbacks, in particular the limited number of molecular species
included.

In this paper, we make a further step ahead. By interpolating on a
grid of opacity tables properly computed by increasing the
abundances of C and N with respect to the scaled solar value, we
calculate new models for low mass AGB stars. We start with a very
metal poor composition ([Fe/H]$<$-2), because the effect of the
TDU is stronger in this case. Observational constraints for these
models come from the so-called CEMP (Carbon Enhanced Metal Poor)
stars, for which C/Fe ratios even larger than $10^3$ have been
observed. Actually, they are not AGB stars, but unevolved dwarfs
belonging to wide binary systems, which have been polluted by the
wind of an AGB companion (see e.g. \citealp{lu05}). The majority
of these CEMP stars also show huge nitrogen enhancements. A slow
deep circulation, perhaps driven by rotation-induced instabilities
or by the formation of magnetic pipes connecting the base of the
convective envelope to the region where the CN cycle takes place,
is generally considered responsible for the partial conversion of
C into N in low mass AGB stars. After \citealp{wasa}~(1995) this
phenomenon has been called {\it Cool Bottom Process} (CBP).
However, nitrogen enhancements can also be produced in very metal
poor stars by a peculiar thermal pulse occurring at the beginning
of the TP-AGB phase, when the abundances of CNO isotopes in the
envelope are particularly low. As firstly proposed by
\citealp{ho90}~(1990) (see also \citealp{iw04} and
\citealp{stra04}), in such a case the convective zone generated by
this anomalous thermal pulse may extend up to the base of the
H-rich envelope. Protons, captured by convection, are rapidly
mixed within the intershell, where the high temperature and the
large C abundance induce a violent H-burning flash. For a short
time, up to 10$^{42}$ erg/sec are released by the CN cycle. Thus,
in the intershell zone, a substantial amount of $^{14}$N (and
$^{13}$C) is produced. Later on, after a particularly deep TDU,
the envelope is enriched with both C and N. If C/O$>$1, a large N
abundance, in addition to the C enhancement, favors the formation
of CN molecules, thus inducing a further increase of the radiative
opacities in the cool atmospheres of AGB stars. In Section 2 we
discuss in more detail the new opacity tables. New models for low
mass AGB stars undergoing the TDU are presented in Section 3,
while in  Section 4 we compare models with only carbon dredge up
and models where both C and N are enhanced. A final discussion
follows.

\section{Molecular opacities}

In the cool layers (\idest temperatures lower than about $4000$ K)
of the convective envelope and of the atmosphere of late-type
stars, molecules become the dominant opacity source. Beside the
local thermodynamic conditions, the concentration of the various
molecular species basically depends on the atomic abundances. In
this respect, an  important quantity is the carbon to oxygen
number ratio. Among the various molecular species involving C
atoms, CO have indeed the larger dissociation energy, so that for
C/O$<1$ almost all the C atoms participate to the formation of
this molecule, while the oxygen atoms in excess are free to form
other molecules. For M stars, the C/O ratio is below unity, \idest
they are "oxygen-rich". During the TP-AGB phase these stars can be
turned, depending on their mass and metallicity, into S-type
(C/O$\simeq$1) and subsequently into carbon-rich objects (N stars
with C/O$>$1), as a consequence of the TDU. In this case, the
carbon atoms in excess  forms new molecules. The atmospheric
structure of an O-rich stars does significantly differ from that
of a C-rich star, as different molecules contribute to the
opacity: TiO and H$_2$O are most important in the oxygen-rich
regime, while carbon-bearing molecules (\eg C$_2$, CN, C$_2$H$_2$
and C$_3$) dominate the opacity for C/O$>$1.

Current model atmosphere codes (\eg hydrostatic MARCS models
\cite{gu75} or dynamic models \citep{ho03}) generally include
molecular opacity data that are suitable for M-type stars as well
as for C-type stars. On the contrary, most stellar evolution
models of AGB stars ignore the opacity changes caused by the TDU.
Recently, the OPAL
collaboration\footnote{http://www-phys.llnl.gov/Research/OPAL/opal.html~.}
as well as the Opacity
Project\footnote{http://opacities.osc.edu/~.} provided web tools
to calculate Rosseland mean opacities. Unfortunately, they are
limited to the atomic contributions and do not include the most
important molecular photon absorbers (except H$_2$). Below 10$^4$
K, the Rosseland mean tables presented by \citealp{ag94}~(1994),
which are available for scaled solar compositions only, are widely
used.

In order to improve our AGB models and to evaluate the effects of
the chemical modifications caused by the TDU and, eventually, by
the CBP, we have calculated a grid of opacity tables by means of
the COMA code from \citealp{Ari2000}~(2000), based on molecular
data given in Tab.~\ref{tab:molecules}. Atomic opacities have been
derived from VALD \cite{kup}. We have generated a set of tables
for different mass fractions of hydrogen, helium, carbon and
nitrogen. The abundances of all the other elements have been
scaled with respect to the corresponding solar values, namely
$X_*=X_\odot \times Z_*/Z_\odot$, where $Z_*=1\times$10$^{-4}$ and
$X_*$ refer to the stellar model composition. As reference solar
composition we adopt \citealp{ag89}~(1989), with some exceptions
(C and N from \citealp{ap02}, while O, Ne and Ar from
\citealp{asp04}). In order to cover the large overabundances
observed in CEMP stars, we calculate tables where C and N have
been independently multiplied up to a factor of 2000 (with
intermediate enhancements of 10, 100 and 500), taking into account
all the possible combinations. Temperatures has been varied to
cover the parameter space of the cool external layers of an AGB
star ($3.2 < \log T <4.05$). We have used the OPAL web tool to
extend the opacity tables at temperature larger than $\log
T=4.05$. Cubic interpolations in $\log T$ and $\log R$
\footnote{$R=\rho \times 10^{18}/T^3$.} and linear interpolations
in the mass fractions of H, C and N ensure a sufficient accuracy
in the evaluation of the opacity coefficients. We decide to first
compute C- and N-enhanced opacity tables at $Z=1\times$10$^{-4}$
because the effects induced by the carbon dredged up in the
envelope is maximized at low metallicity. However, we plan to
prepare C- and N-enhanced tables covering a larger metallicity
range and to collect them in a database.

\begin{figure*}
\resizebox{\hsize}{!}{\plotone{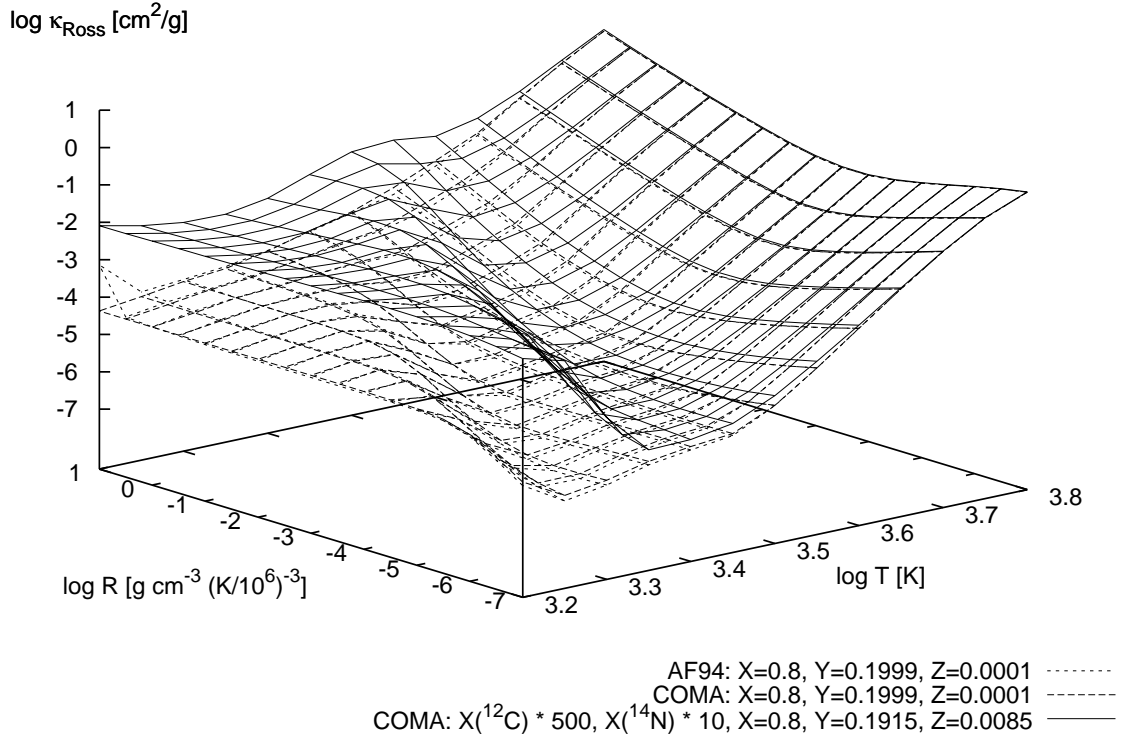}}
\caption{Rosseland mean opacity as a function of $\log T$ and
$\log R$ in a range representative for the envelopes of AGB stars:
dashed lines represent values for a chemical composition X=0.8,
Y=0.1999, and Z=0.0001. As a comparison, the results of
\citealp{ag94}~(1994), which are based on a different line data
set, are shown (dotted lines); the strong discrepancy at low $\log
T$ and high $\log R$ is due to grain opacity which is not included
in our calculations. Enhancing the $^{12}$C (and $^{14}$N) mass
fraction results in a significant increase of opacity (solid
lines) in the cooler layers due to the favored formation of
carbon-bearing molecules, especially CN and
C$_2$H$_2$.\label{fig:lederer1}}
\end{figure*}

\begin{figure*}
\resizebox{\hsize}{!}{\plotone{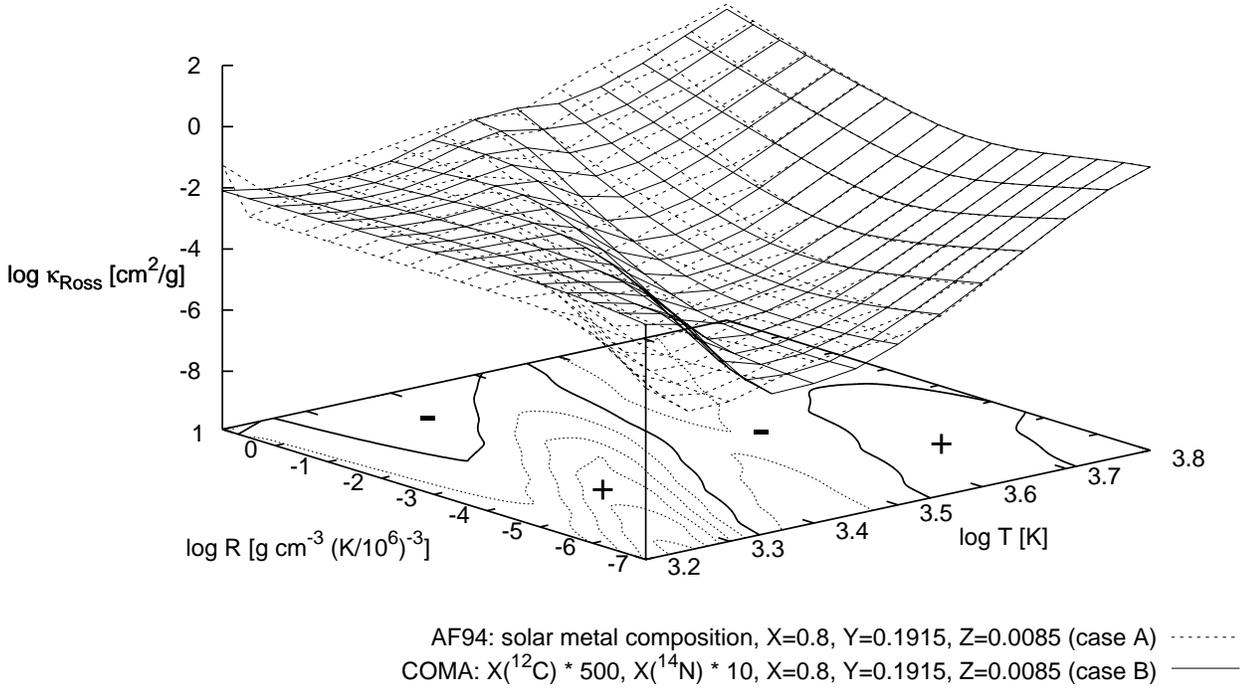}} \caption{ Comparison of
the Rosseland mean opacity for two chemical mixtures with the same
overall metallicity Z=0.0085. Case A is a mixture with solar
scaled metal abundances (dashed lines). For case B only the mass
fractions of $^{12}$C and $^{14}$N have been enhanced starting
from a solar scaled mixture with Z=0.0001 (solid lines). The
differences between the two tables are indicated by contour lines
at the base of the plot. The thick line, at which in terms of
opacity B $=$ A, separates regions where B $>$ A and B $<$ A,
marked with $+$ and $-$ signs, respectively (dotted lines are
separated from each other by steps of 0.5 dex). For case A, H$_2$O
and TiO dominate $\kappa_{\textrm{\scriptsize Ross}}$ at lower
temperatures. At high densities, this leads to a higher mean
opacity than in case B, whereas at low densities carbon-bearing
molecules (cf. Fig.~\ref{fig:lederer2}) cause an increase of the
mean opacity of up to 2.5 dex compared to case A. Atomic
opacities, relevant from $\log T \simeq$ 3.3 to $\log T \simeq$
3.8 (depending on $\log R$), are higher in case A.
\label{fig:lederer1a}}
\end{figure*}

\begin{figure*}
\resizebox{\hsize}{!}{\includegraphics[angle=90,scale=1.0]{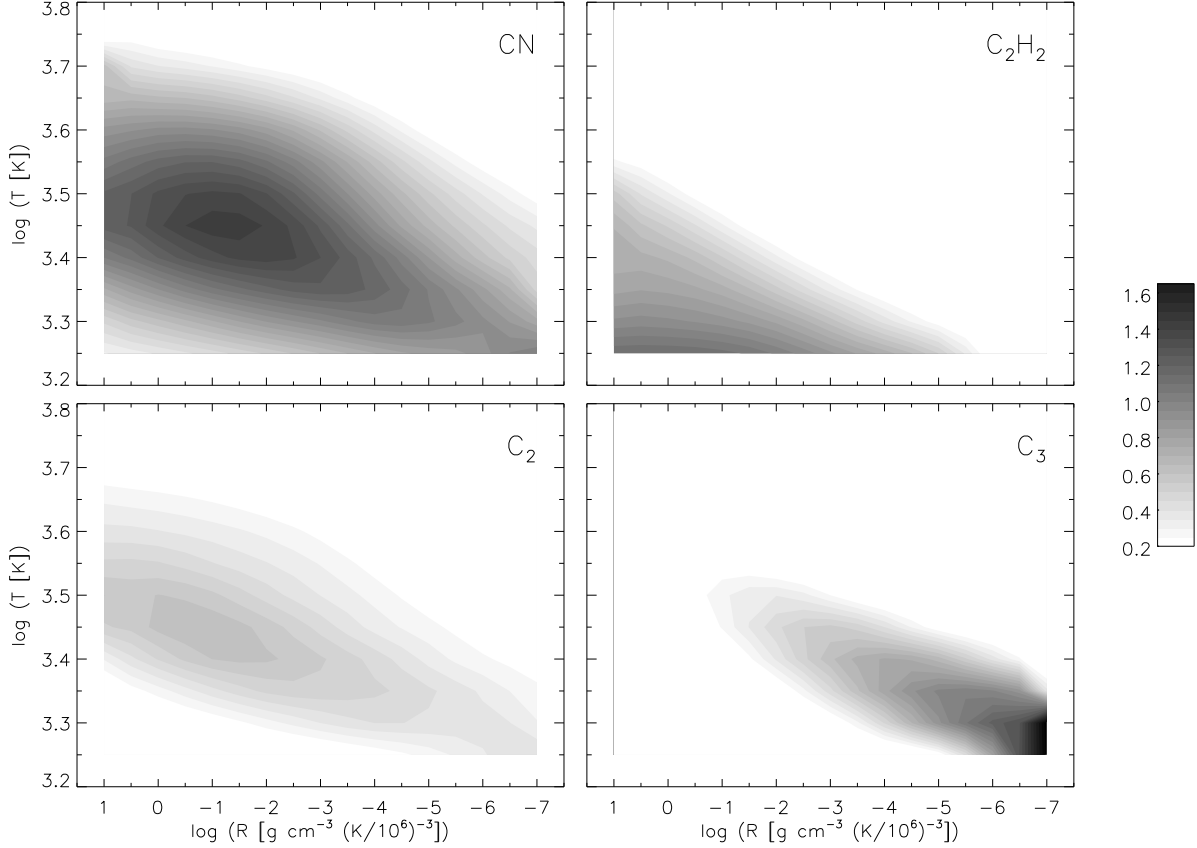}}
\caption{Contribution of molecular species to the Rosseland mean
opacity: the plots show differences (in orders of magnitudes)
between the total opacity and calculations where particular
molecules (indicated in each panel) have been omitted. The extreme
case of our grid (\idest mass fraction of $^{12}$C and $^{14}$N
enhanced by a factor of 2000) has been chosen as the basis for
these figures, as in this way the regions where certain molecules
contribute can be indentified most clearly.\label{fig:lederer2}}
\end{figure*}

The differences between opacities calculated for scaled solar
composition and opacities calculated by enhancing C and N are
illustrated in Figs.~\ref{fig:lederer1} and \ref{fig:lederer1a}.
As shown in Fig.~\ref{fig:lederer1}, an enhancement of carbon and
nitrogen due to the third dredge up leads to a C- (and N-) rich
mixture whose opacities are more than two orders of magnitude
larger than the initial ones over a wide range of temperature and
density. In this Figure, we also compare our initial scaled solar
opacity to the corresponding values from \citealp{ag94}~(1994).
The overall agreement is quite good. The significant discrepancy
at low temperature/high density is due to the inclusion of dust
absorption in \citealp{ag94}~(1994). Even though the stellar
models here considered never attain the physical conditions for
dust grain formation in their atmospheres, we want to point out
that, for the astrophysical scenario under consideration, dust
formation is far from equilibrium; usually much less dust than
under equilibrium conditions is formed \cite{gail}. In
Fig.~\ref{fig:lederer1a} we point out the differences between the
mean opacity coefficients for a scaled solar composition (case A,
derived from \citealp{ag94}~1994) and a corresponding table where
only the C and N abundances are enhanced (case B). Particularly,
in case B we start from a scaled solar mixture with a metallicity
of Z=$10^{-4}$ and then we enhance the mass fraction of $^{12}$C
by a factor of 500 and the mass fraction of $^{14}$N by a factor
of 10, thus attaining a total metallicity Z=8.5$\times 10^{-3}$
(the same as in case A). In the oxygen-rich case A, H$_2$O and TiO
dominate the opacity at low temperatures. For high densities, the
Rosseland mean is higher than in the carbon-rich case B, whereas
the opposite is true for lower densities. In this region
carbon-bearing molecules are the dominant opacity source (see
below and cf. Fig.~\ref{fig:lederer2}). In the region from $\log T
\simeq$ 3.3 to $\log T \simeq$ 3.8 (depending on $\log R$), atomic
opacities significantly contribute to the Rosseland mean.
Consequently, case A comprises higher opacities in this region as
more metals (apart from C and N) are present than in case B.

The most important contributions in the Rosseland mean opacities
of a C-rich mixture are due to CN, C$_2$, C$_2$H$_2$, and C$_3$.
These molecules contribute to the mean opacity in different
regions of the parameter space ($\log T, \log R$) (see
Fig.~\ref{fig:lederer2}). In particular, C$_2$H$_2$ dominates the
low temperature high density region, while C$_3$ provides the main
photon absorbtion at low T and low R, but only for extreme C-rich
mixture. At higher temperatures (3000$<$T$<$4000) and densities
the opacity is mainly controlled by CN and C$_2$, and the relative
contribution between these two molecules also depends on the
amount of nitrogen. Note that, at variance with solar metallicity,
in metal poor stars, even considering the nitrogen enhancement
resulting after the first dredge up, the CN molecules may
contribute to the opacity only if primary nitrogen is added, as in
the case of many CEMP stars.

Finally, we have also investigated the influence of a moderate
increase of the oxygen mass fraction. Compared to the effects of
the carbon enhancement, this contribution was found to be
negligible. For this reason, variations of the O abundance were
not included into our set of tables.

\section{C-enhanced models}\label{sec:newmodels}

As outlined in the previous section, molecules are the dominant
opacity source in the range 3.3 $<$ logT $<$ 3.6. These
temperatures are usually attained in the most external layers of
AGB stars. As far as we know, the extant AGB stellar models are
based on low temperature opacity tables calculated for scaled
solar compositions and only the variation of the relative
abundance of H and He is generally taken into account. Perhaps,
some investigators could have approximated the effect of
dredged-up by using opacities appropriate to the new total Z
value. However, while this approximation may be adequate to follow
the evolution of AGB progenitors, the occurrence of recursive TDU
episodes and the consequent enhancement of C in the envelope
requires particular attention. The inadequacy of scaled solar
opacity coefficients for AGB stellar models is particularly severe
in metal poor calculations, where the overall mass of C that is
dredged up may be 1000 times larger than the whole initial metal
content (Z), whereas the iron content practically remains equal to
the initial one.

In our previous calculations (\citealp{cri06}~2006,
\citealp{stra06}~2006), attempting to mimic the effect of the C
enhancement  on the opacity, we choose to linearly interpolate on
Z, the overall mass fraction of elements with atomic number $\geq
6$. In that case, we didn't allow for possible relative variations
among these elements. Actually, the material that is dredged up by
convection is mainly composed by freshly synthesized C, while
other heavy elements, as iron, remain practically unchanged. Thus,
such a global metallicity interpolation overestimates the atomic
iron contribution to the opacity, whereas underestimates the
carbon (atomic and molecular) contribution. Since these different
atomic and molecular opacity sources operates in distinct stellar
layers, it is rather difficult to evaluate, a priori, how much the
contributions from these two different sources of opacity may
compensate each other.

In this Section we compare stellar models with initial mass
$M$=2$M_\odot$, initial helium Y=0.245 and initial metallicity
$Z$=1$\times$10$^{-4}$ (or [Fe/H]=-2.17), as obtained under
different assumptions about opacity. To calculate the
corresponding evolutionary sequences, we have used the  FRANEC
stellar evolution code (the release described in \citealp{chi98}).
Mixing and mass loss algorithms for the AGB phase are those
described in \citealp{stra06}~(2006). In particular, we derive the
mass loss rate for AGB models basing on the observed correlation
with the pulsational period (\citealp{scol},\citealp{wh03}).
Periods are estimated from the P-M$_{\rm K}$ relation proposed by
\citealp{feast}~(1989)\footnote{Bolometric corrections from
\citealp{flu94}~(1994) have been used.}.

We discuss, in particular, three different models, namely:

\begin{enumerate}
\item{the {\it Z-fixed} model, where only variations of He/H are
considered (the metallicity is always maintained equal to the
initial value);} \item{the {\it Z-int} model, where variations of
He/H and Z are considered, but the elemental distribution of heavy
elements is always maintained scaled solar;} \item{the {\it
CN-int} model, which properly takes into account variations of
He/H, C and, eventually, N caused by the TDU. All the other
elements are maintained to their initial (scaled solar) values.}
\end{enumerate}

Let us remark that the three models have been computed with
identical input physics, except the opacity coefficients, and that
the three evolutionary sequences have been followed up to the last
TDU episode.

\begin{figure*}
\resizebox{\hsize}{!}{\includegraphics{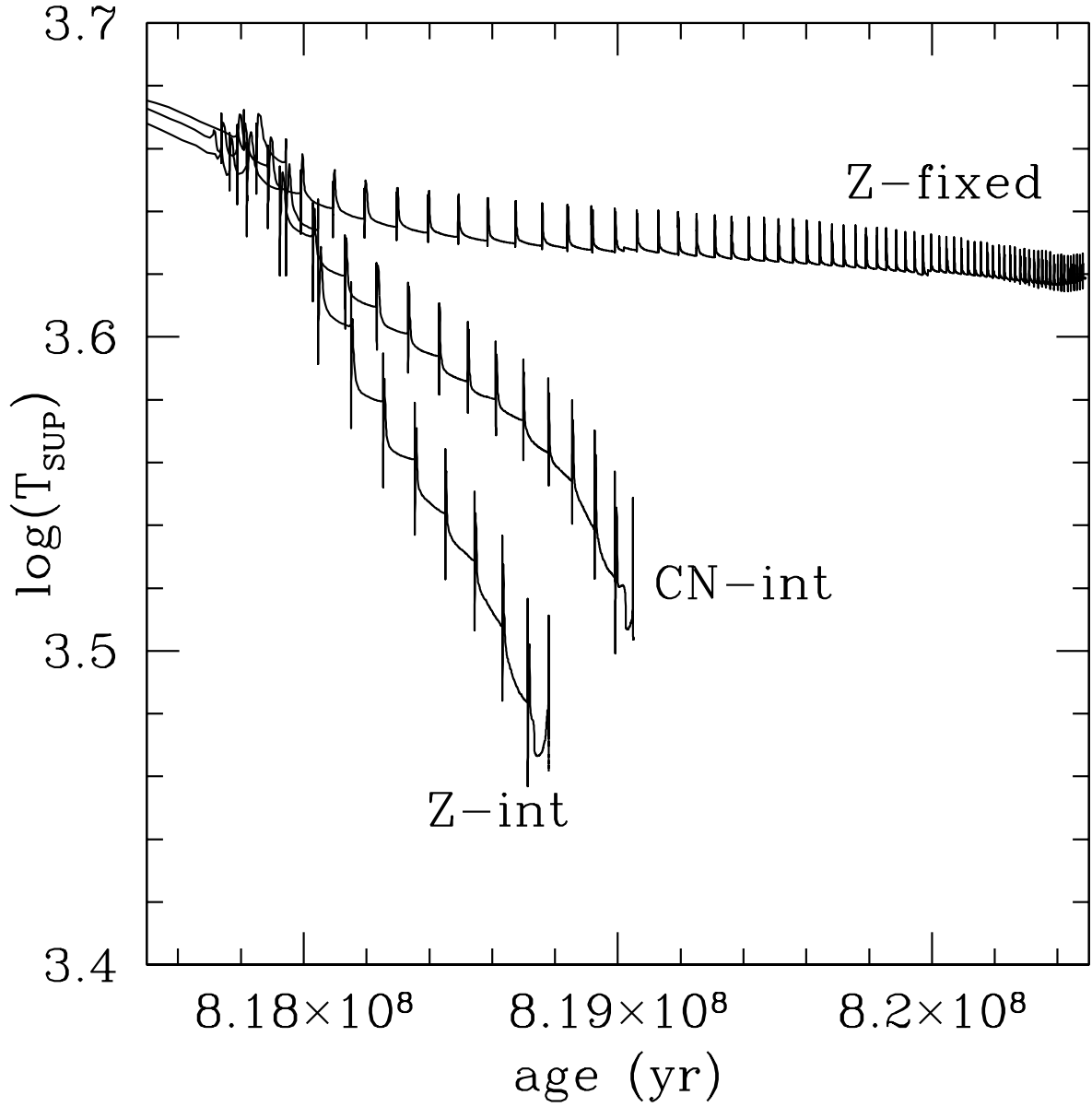}} \caption{We
report, for the three models discussed in the text, the surface
temperature vs. age.} \label{fig3}
\end{figure*}

Fig.~\ref{fig3} clearly shows the huge difference in AGB evolution
resulting from different opacity calculations. The first direct
consequence of the introduction of a growing envelope opacity in
the {\it Z-int} and {\it CN-int} models is the evident drop of the
effective temperature. The temperature gradient is indeed linearly
dependent on the opacity coefficient. For a given luminosity, this
occurrence implies a larger radius as well as a larger mass loss
rate. As a result, the duration of the TP-AGB phase is
significantly shortened. This effect is however overestimated in
the {\it Z-int} model with respect to the more appropriate {\it
CN-int} model.

\begin{figure*}
\resizebox{\hsize}{!}{\includegraphics{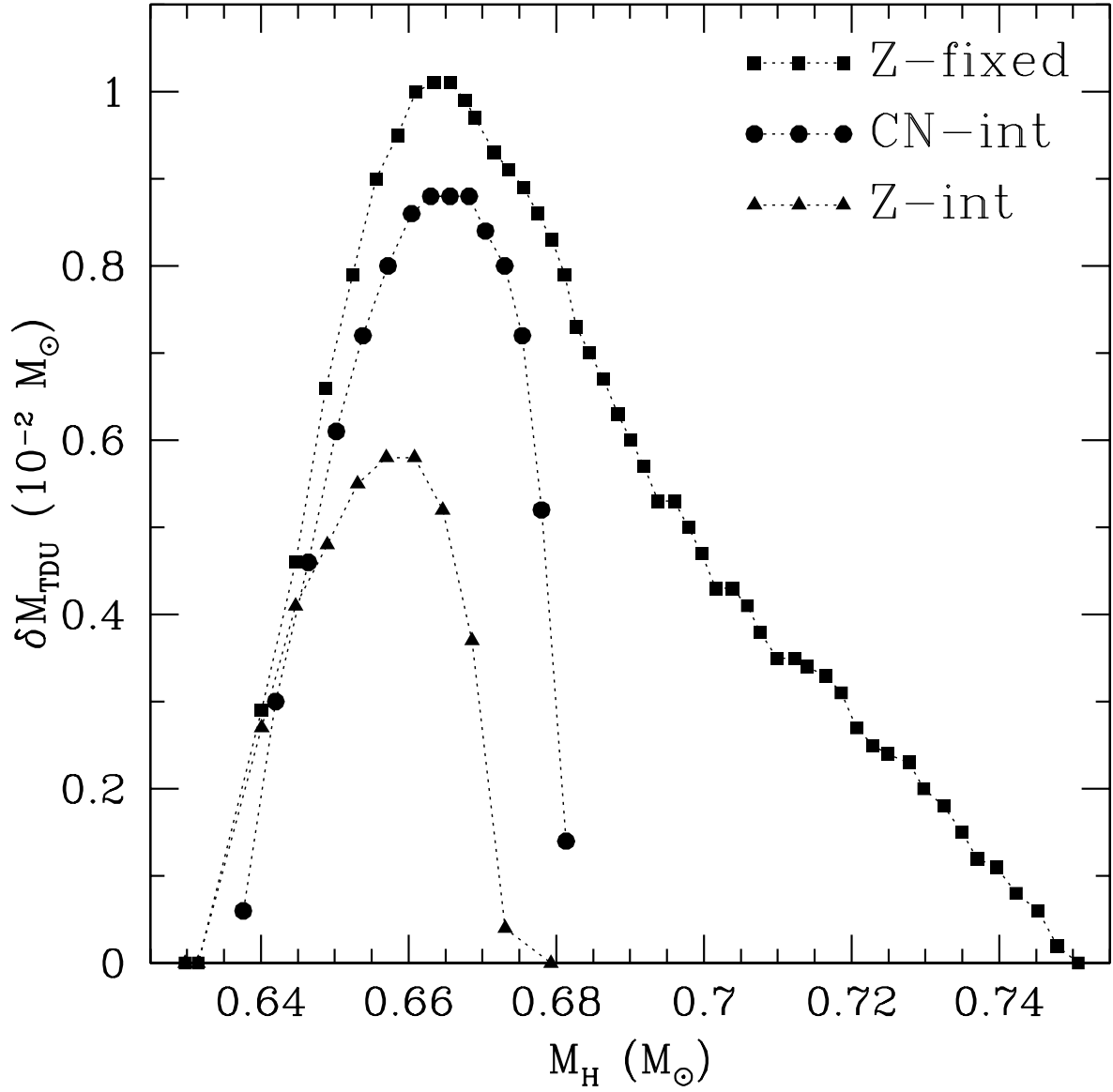}} \caption{We
report, for the three models discussed in the text, the dredged up
material vs. core mass.} \label{fig32}
\end{figure*}

A second important consequence of the different treatment of the
low  temperature opacity concerns the chemical yields. The amount
of mass that is dredged up after a thermal pulse ($\delta
M_{TDU}$) basically depends on the core mass and on the envelope
mass and both of them are affected by a variation of the envelope
opacity. The evolution of the three calculated $\delta M_{TDU}$
are illustrated in Fig.~\ref{fig32}. They initially increase,
because the core mass increases, reach a maximum and, then,
decrease, because the mass loss erodes the envelope mass. However,
since the growth of the mass loss is steeper in the {\it Z-int}
and {\it CN-int} models, the total mass that is dredged up is
lower than that found in the case of the {\it Z-fixed} model. As a
result, models computed with larger opacity predict a smaller
contribution to the galactic chemical evolution.

\citealp{cri06} (2006) already noted that the overabundances of
the heavy elements synthesized by the s process in the {\it Z-int}
model are generally too small compared to the  available
spectroscopic determination for C- and s-rich metal poor stars
belonging to the galactic halo. Such a discrepancy appears more
severe, considered that these halo stars are not AGB stars
undergoing the TDU, but less evolved objects, dwarfs or red
giants, whose envelopes have been polluted by the wind of already
evolved AGB companions. After the rapid accretion process, the C-
and s- enriched material has been diluted by convection, in case
of giant stars, or by secular processes, as microscopic diffusion
or thermohaline mixing, in case of dwarf stars (see e.g.
\citealp{sta07}). We concluded that the opacity is overestimated
in these models, causing a too large mass loss and a too short
TP-AGB phase.

A longer TP-AGB phase, with more dredge up episodes and, in turn,
with larger final overabundances of the s-elements, are obtained
in the case of  the {\it Z-fixed} model, but in this case the
effective temperature, always greater than 4200 K (see the upper
curve in Fig.~\ref{fig3}), would not match the observed
temperatures of metal poor C(N) stars found in dwarfs spheroidal
galaxies, typically ranging between 3500 and 4000 K
(\citealp{dom04}~2004).

An inspection of Fig.~\ref{fig32} shows that the {\it CN-int}
model, which undergoes 15 TDU episodes, represents an intermediate
case between the {\it Z-int} model (9 TDU) and the {\it Z-fixed}
model (48 TDU). The total amount of mass that is dredged up is
3.8$\times 10^{-2}$ M$_\odot$, 9.7$\times 10^{-2}$ M$_\odot$ and
1.9$\times 10^{-1}$ M$_\odot$ in the {\it Z-int}, {\it CN-int} and
{\it Z-fixed} model, respectively. Even if the {\it CN-int} model
shows lower overabundances with respect the {\it Z-fixed} model,
the drop of its effective temperature caused by the C dredge up
matches the low values found for metal poor C(N) stars in dwarf
spheroidal galaxies. It appears, therefore, that the inconsistency
between surface temperatures and s-process enhancement could be
solved by using appropriate opacity coefficients. A detailed
comparison between the prediction of our AGB models and the
observed elemental overabundances in CEMP stars will be presented
in a forthcoming paper.

\section{C- and N-enhanced models} \label{cpb}

\begin{figure*}
\resizebox{\hsize}{!}{\includegraphics{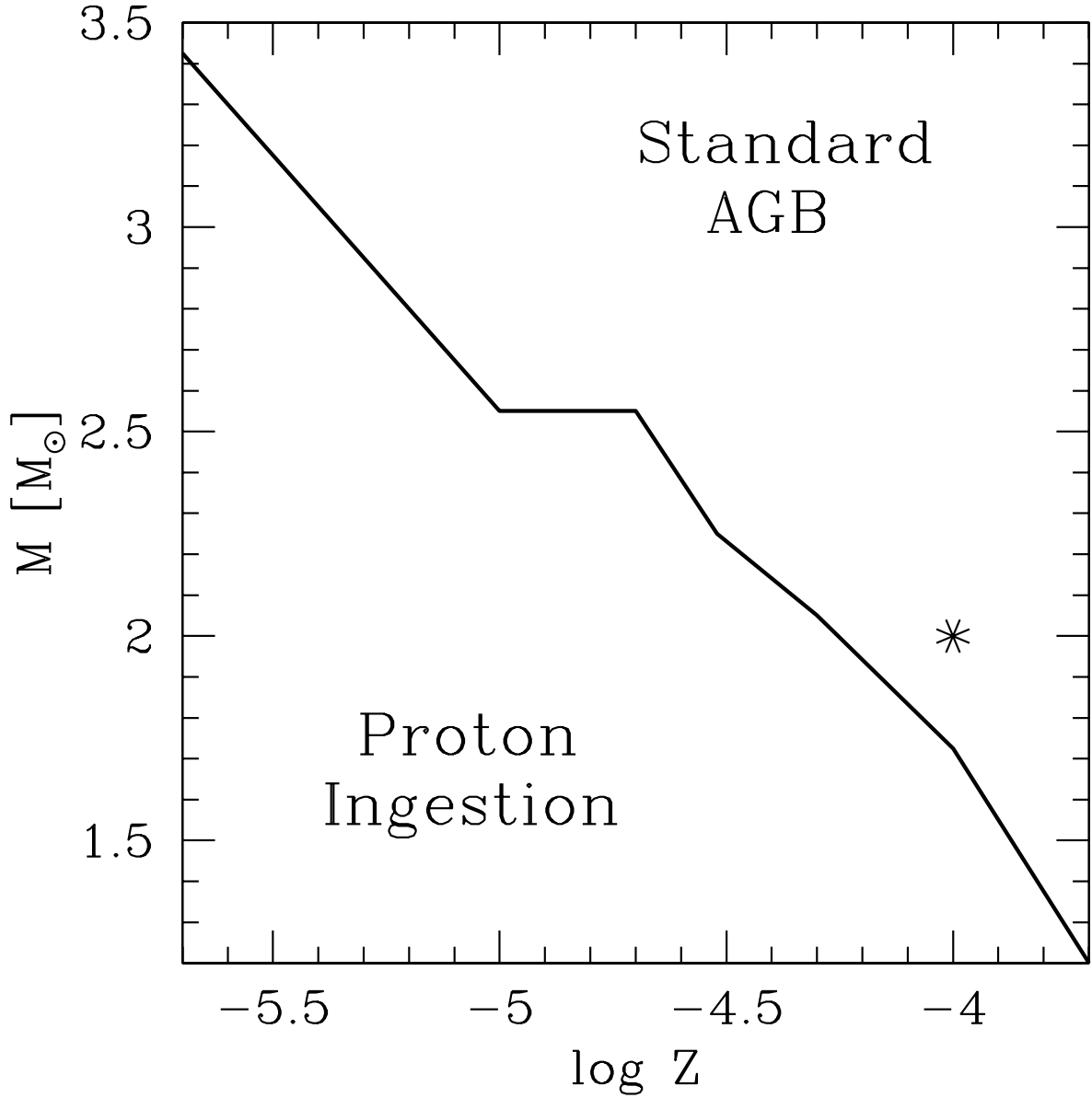}} \caption{In a
metallicity-mass diagram, we report the line that separates models
following a standard AGB evolution (upper region) from that ones
suffering a proton ingestion episode at the beginning of their AGB
phase (lower region). The asterisk corresponds to the initial
input parameters of the models presented in this
paper.}\label{fig4}
\end{figure*}

In massive AGB (M$>4$ M$_\odot$) the temperature at the base of
the convective envelope is large enough to convert part of the C
dredged up into N (the so called Hot Bottom Burning). This is not
the case of low mass AGB, although some authors postulated the
existence of a slow circulation, named Cool Bottom Process (CBP),
capable to mix the material within the thin layer located between
the H-burning shell and the cool base of the convective envelope
\cite[see][and references therein]{no03}. While it is widely
accepted that the CBP operates during the RGB, mainly because low
mass red giant stars show low values of the $^{12}$C/$^{13}$C
ratio \cite{gra00}, its activation during the AGB still remains a
matter of debate. Up to date, the origin of CBP has not been
clearly identified, but the magnetic field could play a relevant
role for the activation of this process \citep{bu06}. If the CBP
would be at work in low mass AGB, it might be responsible for the
conversion of C into N. Actually, many CEMP stars show large N
enhancements. However, as already recalled, a huge N enhancement
might be also produced if the convective zone generated by a
thermal pulse ingests protons from the top, a possibility early
recognized by \citealp{ho90}~(1990), who claimed that this process
should be rather common in very metal poor stars. We are preparing
a paper where we describe the evolution and the nucleosynthesis of
these models. Preliminary results indicate that it exists a
maximum mass for which such a peculiar thermal pulse takes place
and that this limit increases as the metallicity decreases. In
Fig.~\ref{fig4}, the asterisk represents the initial input
parameters of the models presented here.

\begin{figure*}
\resizebox{\hsize}{!}{\includegraphics{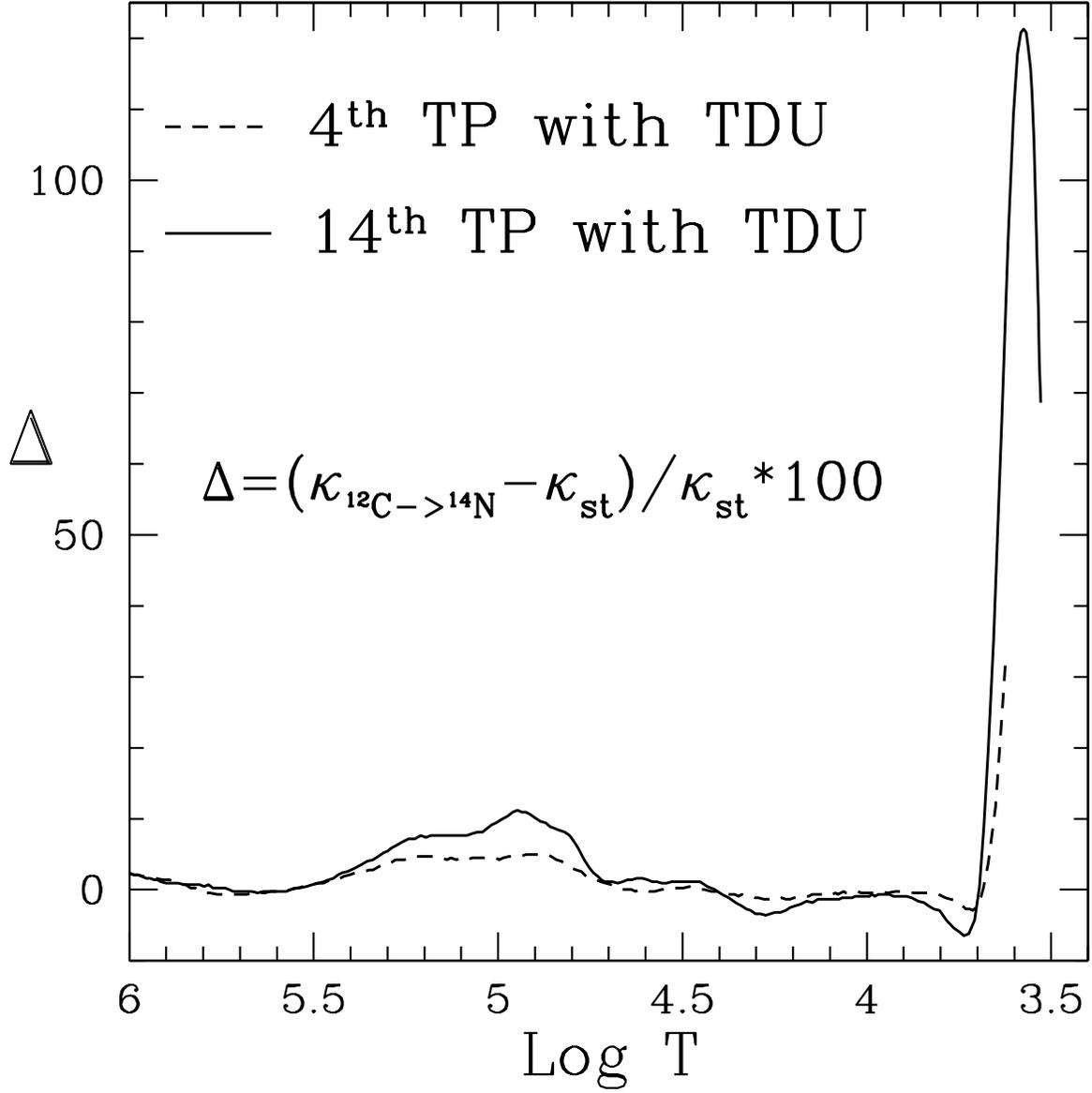}}
\caption{Percentage differences of the opacity coefficients
calculated in the physical structure of the {\it CN-int} model
after the 4$^{th}$ TP with TDU (dashed line) and after the
14$^{th}$ TP with TDU (solid line); see text for details.}
\label{fig5}
\end{figure*}

Our interest, in the context of this paper, is related to the
possible consequences induced by the N enhancement on the
properties of low-mass low-metallicity AGB stars. A first
indication about the difference between a C-rich and a CN-rich
envelope can be deduced from Fig.~\ref{fig5}. Here, starting from
the physical structure of the {\it CN-int} model after the
4$^{th}$ and the 14$^{th}$ pulse with TDU, we calculate, without a
re-adjustment of the stellar structure, the opacity coefficients
assuming that one half of the $^{12}$C has been converted into
$^{14}$N. The resulting percentage differences with respect to the
original opacity of the {\it CN-int} model (those calculated
taking into account the C dredge up only) have been reported.

The important differences found at low temperature are due to the
absorbtion of the CN molecules, whose contribution to the opacity
is marginal in the {\it CN-int} model. This effect is larger
toward the end of the AGB phase, because the surface temperature
drops down to 3500 K (logT$=$3.55), where the CN contribution is
maximum (see Fig.~\ref{fig:lederer2}).

In order to better characterize the influence of the N
enhancement, four additional AGB models have been computed, by
introducing an extra-circulation below the base of the convective
envelope. As shown by \citealp{dom04}~(2004), the most important
parameter of the CBP is the maximum temperature at which the
circulated material is exposed (T$_{max}$), while the actual rate
of circulation is significantly less important. Then, in the four
models here presented, the circulation rate is taken constant and
fixed to 1/1000 of the average turbulent velocity of the most
internal layers within the convective envelope ( corresponding to
$v_{conv}\sim (20\div 100)\; {\rm cm} \cdot {\rm s}^{-1}$), while
the mass extension of the extra-circulation is varied so that
T$_{max}$=30, 40, 50 and 60$\times 10^6$ K in model {\it
CN-int-30}, {\it CN-int-40}, {\it CN-int-50} and {\it CN-int-60},
respectively. The extra-circulation has been switched on at the
beginning of the TP-AGB phase and it remains active only if the
energy flux released by the H-burning is larger than that of the
He-burning({\it i.e.} during the interpulse phase).

When T$_{max} \ge 40 \times 10^6$ K, the larger opacity caused by
the $^{14}$N production from CBP makes the surface temperature
lower with respect to the standard (i.e. {\it CN-int}) model (for
T$_{max}= 30 \times 10^6$ K no appreciable differences have been
found). This implies a more efficient mass loss and a shorter
TP-AGB phase. The {\it CN-int-40} model experience 12 TDU
episodes, 11 the {\it CN-int-50} model and 9 the {\it CN-int-60}
model. The total masses that are dredged up are 7.6$\times
10^{-2}$ M$_\odot$, 6.6$\times 10^{-2}$ M$_\odot$ and 6.3$\times
10^{-2}$ M$_\odot$ , respectively.

\begin{figure*}
\resizebox{\hsize}{!}{\includegraphics{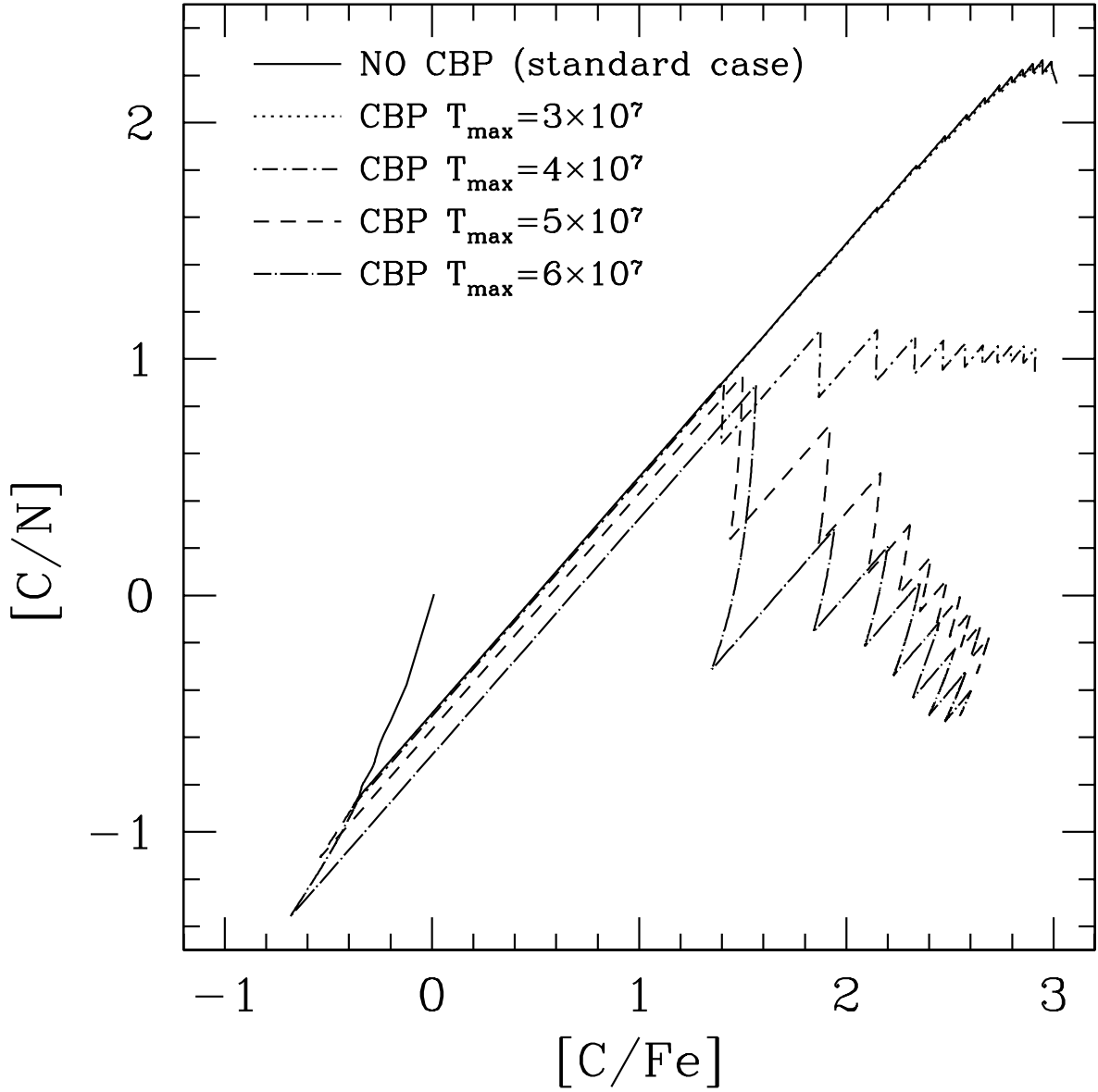}} \caption{[C/N]
surface ratios plotted against [C/Fe] for different CBP
efficiencies. See text for details.} \label{fig52}
\end{figure*}

The envelope chemical composition is significantly affected by the
introduction of the CBP, the surface nitrogen abundance being
greatly enhanced with respect to the standard case. In
Fig.~\ref{fig52}, we report the evolution of the [C/N] ratio
versus the [C/Fe] ratio, in the usual spectroscopic notation, for
the four models with CBP and the one without CBP.

All these models start the pre-main sequence with scaled solar abundance ratios,
namely [C/Fe]=0 and [C/N]=0, whose values drop down to -0.31 and -0.74, respectively,
 after the first dredge up.
Then, the resulting saw-blade pattern of the various curves is due
to the alternate action of the TDU and the CBP during the AGB
phase. Indeed, [C/Fe] and [C/N] both increase after a TDU episode,
whereas they decrease during interpulse periods, as a consequence
of the CBP. The deeper the CBP is, the lower the final [C/N] ratio
is, spanning a range of about two order of magnitude (from
[C/N]$\sim$2.2 down to [C/N]$\sim$-0.5). Note that the [C/N] of
the CEMP stars, for which this ratio has been derived, typically
ranges between 0 and 1 \cite{jo07}.

\section{Conclusions}

In this paper we have discussed the use of appropriate opacity to
describe the effects of  the C enhancement caused by the third
dredge up in low-mass-low-metallicity AGB stars. New opacity
tables for chemical mixtures with different overabundances of C
and N  have been obtained by means of the COMA code
(\citealp{Ari2000}~2000). Then, stellar models with
$M$=2$M_\odot$, Y=0.245 and $Z$=1$\times$10$^{-4}$ have been
computed, varying the interpolation scheme on these tables. We
have also discussed the consequence of the conversion of C into N,
as eventually caused by the Cool Bottom Process operating in low
mass AGB stars or by proton ingestion in the first Thermal Pulse
of very low metallicity AGB stars.

Both the C dredge up and the conversion of C into N induce
substantially affect the molecular contribution to the opacity for
temperature lower than 4000 K. Larger opacity coefficients imply
cooler envelopes, larger mass loss rate and, in turn, shorter AGB
lifetime. It also affects the variation of the surface composition
and the global yields produced by low mass AGB stars. The amount
of mass that is dredged up is, indeed, influenced by the different
temporal variation of the envelope mass.

We show that only models computed by adopting opacity that
properly include the enhancements of C and, eventually, N can
reproduce the photometric and spectroscopic properties of their
observational counterparts. In particular, the low effective
temperature ($3500\sim4000$ K) of the low metallicity C stars
belonging to dwarf spheroidal galaxies can be naturally reproduced
when the new opacity are used. These models can also account for
the relatively large overabundance of heavy elements, those
produced by the s-process nucleosynthesys, observed in C- and
s-rich halo stars.

\acknowledgments

SC and OS have been supported by the Italian National Grant
Program PRIN 2004. MTL has been supported by the Austrian Academy
of Sciences (DOC-programme) and acknowledges funding through FWF
project P18171.

\clearpage

\begin{deluxetable}{lcrc}
\tablecaption{Molecular Line Data\label{tab:molecules}}
\tablewidth{0pt}
\tablehead{
\colhead{Molecule} & \colhead{Source of Thermodynamic Data} & \colhead{Number of Lines} & \colhead{Source of Line Data}
}
\startdata
CO         & 1 &    131,859 &  6 \\
CH         & 2 &    229,134 &  7 \\
C$_2$      & 1 &    360,887 &  8 \\
SiO        & 2 &     93,372 &  9 \\
CN         & 1 &  2,533,040 &  7 \\
TiO        & 1 & 22,758,691 & 10 \\
H$_2$O     & 3 & 27,988,952 & 11 \\
HCN/HNC    & 4 & 34,433,190 & 12 \\
OH         & 1 &     38,068 & 13 \\
VO         & 1 &  3,171,552 & 14 \\
CO$_2$     & 2 &     60,802 & 15 \\
SO$_2$     & 5 &     38,853 & 15 \\
HF         & 1 &        107 & 15 \\
HCl        & 1 &        533 & 15 \\
\hline
C$_2$H$_2$ & - & opacity sampling & 16 \\
C$_3$      & - & opacity sampling & 16 \\
\enddata
\tablecomments{Molecules entering the calculation of the Rosseland
mean opacity. References for thermodynamic and line data as
indexed in columns two and four are given below. Atomic line data
are taken from VALD \cite{kup}. } \tablerefs{ (1) \citep*{st84};
(2) \citep*{ro85}; (3) \citep{vt00}; (4) \citep*{ba02}; (5)
\citep*{ir88}; (6) \citep*{gc94}; (7) \citep*{jo97}; (8)
\citep*{qu74}; (9) \citep*{la93}; (10) \citep*{sc98}; (11)
\citep*{ba06}; (12) \citep*{ha06}; (13) \citep*{sc97}; (14)
\citep*{ap98}; (15) \citep*{ro98}; (16) \citep*{jo89}.}
\end{deluxetable}

\clearpage

\clearpage

\clearpage

\clearpage

\end{document}